\documentclass[aps,
	twocolumn,
	altaffilletter,
	nolongbibliography,
	numerical,
	flushbottom,
	secnumarabic,
	prl,
	superscriptaddress,
	floatfix,
	10pt]{revtex4-2}

\usepackage{graphicx}
\usepackage{dcolumn}
\usepackage{bm}%
\usepackage{changepage}
\usepackage{orcidlink}
\usepackage{newtxtext,newtxmath}
\usepackage{dsfont}
\usepackage{comment}
\usepackage[toc,page]{appendix} 
\usepackage{hyperref}
\hypersetup{%
	colorlinks=true, pdfstartpage=1, pdfstartview=FitH, pdfborder={0 0 0},%
	breaklinks=true, pdfpagemode=UseNone, pageanchor=true, pdfpagemode=UseOutlines,%
	plainpages=false, bookmarksnumbered, bookmarksopen=true, bookmarksopenlevel=1,%
	hypertexnames=true, pdfhighlight=/O,
	urlcolor=blue, linkcolor=blue, citecolor=blue
}
\usepackage{graphicx}

\usepackage[normalem]{ulem}
\usepackage{soul}

\newcommand{\bra}[1]{\langle #1 |}
\newcommand{\ket}[1]{| #1 \rangle}
\newcommand{\be}{\begin{equation}}
\newcommand{\ee}{\end{equation}}
\newcommand{\beN}{\begin{equation*}}
\newcommand{\eeN}{\end{equation*}}
\newcommand{\ba}{\begin{eqnarray}} 
\newcommand{\ea}{\end{eqnarray}}   
\newcommand{\id}{\mathds{1}}
\DeclareMathOperator{\tr}{Tr}

\newcommand{\Ham}{\hat{\mathcal{H}}}
\newcommand{\Hamaux}{\hat{H}_\mathrm{aux}}
\newcommand{\Projector}{\hat{\Pi}}

\newcommand{\ud}{\mathrm{d}}
\newcommand{\PauliSigma}{\hat{\sigma}}
\newcommand{\calL}{\mathcal{L}}
\newcommand{\calP}{\mathcal{P}}

\newcommand{\calN}{\mathcal{N}}
\newcommand{\opA}{\hat{A}}
\newcommand{\opB}{\hat{B}}
\newcommand{\opL}{\hat{L}}
\newcommand{\opP}{\hat{P}}
\newcommand{\identity}{{\mathds 1}}
\newcommand{\Project}{{\mathds P}}
\newcommand{\rmj}{\mathrm{j}}
\newcommand{\defuguale}{\stackrel{\textrm{def}}{=}}
\newcommand{\Unitary}{\hat{U}}
\newcommand{\opc}[1]{\hat{c}^{\phantom \dagger}_{#1}}
\newcommand{\opcdag}[1]{\hat{c}^{\dagger}_{#1}}

\newcommand{\ignore}[1]{}

\begin{document}

\preprint{APS/123-QED}

\title{Parent Hamiltonian reconstruction via inverse quantum annealing}

\author{Davide Rattacaso~\orcidlink{0000-0001-8219-5806}}
\email{davide.rattacaso@pd.infn.it}
\affiliation{Dipartimento di Fisica ``E. Pancini", Universit\`a di Napoli ``Federico II'', Monte S. Angelo, I-80126 Napoli, Italy}
\affiliation{Dipartimento di Fisica e Astronomia ``G. Galilei",
Universit\`a di Padova, I-35131 Padova, Italy}
\author{Gianluca Passarelli~\orcidlink{0000-0002-3292-0034}}
\affiliation{CNR-SPIN, c/o Complesso di Monte S. Angelo, via Cinthia - 80126 - Napoli, Italy}
\author{Angelo Russomanno~\orcidlink{0009-0000-1923-370X}}
\affiliation{Scuola Superiore Meridionale, Università di Napoli Federico II, Largo San Marcellino 10, I-80138 Napoli, Italy}
\affiliation{Dipartimento di Fisica ``E. Pancini", Universit\`a di Napoli ``Federico II'', Monte S. Angelo, I-80126 Napoli, Italy}
\author{Procolo Lucignano~\orcidlink{0000-0003-2784-8485}}
\affiliation{Dipartimento di Fisica ``E. Pancini", Universit\`a di Napoli ``Federico II'', Monte S. Angelo, I-80126 Napoli, Italy}
\author{Giuseppe E. Santoro~\orcidlink{0000-0001-6854-4512}}
\affiliation{SISSA, Via Bonomea 265, I-34136 Trieste, Italy}
\affiliation{The Abdus Salam International Center for Theoretical Physics, Strada Costiera 11, 34151 Trieste, Italy}
\affiliation{CNR-IOM Democritos National Simulation Center, Via Bonomea 265, I-34136 Trieste, Italy}
\author{Rosario Fazio~\orcidlink{0000-0002-7793-179X}}
\affiliation{The Abdus Salam International Center for Theoretical Physics, Strada Costiera 11, 34151 Trieste, Italy}
\affiliation{Dipartimento di Fisica ``E. Pancini", Universit\`a di Napoli ``Federico II'', Monte S. Angelo, I-80126 Napoli, Italy}

\date{\today} 

\begin{abstract}
Finding a local Hamiltonian $\Ham$ having a given many-body wavefunction $|\psi\rangle$
as its ground state, i.e. a \textit{parent Hamiltonian}, is a challenge of fundamental importance in quantum technologies. 
Here we introduce a numerical method, inspired by quantum annealing, that efficiently performs this task through an artificial inverse dynamics: a slow deformation of the states $|\psi(\lambda(t))\rangle$,
starting from a simple state $|\psi_0\rangle$ with a known $\Ham_0$, 
generates an adiabatic evolution of the corresponding Hamiltonian. 
We name this approach \emph{inverse quantum annealing}. 
The method, implemented through a projection onto a set of local operators, only requires the knowledge of local expectation values, and, for long annealing times, leads to an approximate parent Hamiltonian whose degree of locality 
depends on the correlations built up by the states $|\psi(\lambda)\rangle$.
We illustrate the method on two paradigmatic models: the Kitaev fermionic chain and a quantum Ising chain in longitudinal and transverse fields.
\end{abstract}

\maketitle

{\em Introduction.---} The success of quantum technologies ultimately relies on our ability to control increasingly complex artificial quantum systems~\cite{Preskill2018}.  
This may require, in quantum simulators, the accurate tailoring of a many-body Hamiltonian. Controlling~\cite{opt_c_6,opt_c__} and verifying~\cite{verification_2,verification_4} the actual functioning of these systems have raised increasing attention to the search for \emph{parent Hamiltonians} (PHs)~\cite{PH_00, PH_0,PH1,PH2,PhysRevA.86.022339,ff_for_mps0,ff_for_mps,bisognano_1, bisognano_2,Chertkov2018}. This problem consists in finding 
a local and/or engineerable Hamiltonian having a given wavefunction as a ground state. The knowledge of a PH is related to Hamiltonian learning~\cite{h_learning_0,h_learning_1,Granade_2012} and verification of quantum devices, and can be exploited to experimentally prepare a target ground state. The search for a PH represents an especially complex instance of the reconstruction of a Hamiltonian from one of its eigenstates~\cite{Qi2019,Greiter2018,Bairey2019,Hou2020,Cao2020} or time-dependent states~\cite{PhysRevA.103.052403, PhysRevResearch.3.023246, bairey2021,Rattacaso2021,Franca2022,RATTACASO2022_HIGH}. In particular, the space of the Hamiltonians having a given state as an eigenstate can be efficiently reconstructed from correlation functions~\cite{Qi2019,Chertkov2018,Greiter2018} or expectation values of local commutators~\cite{Bairey2019}. Picking PHs in this 
space is a hard task since it generally requires the diagonalization of all the candidate PHs to verify that the target state is a ground state~\cite{Chertkov2018}. 
More efficient methods, based on local measurements, have been suggested to obtain approximate PHs~\cite{Xin2019, Hou2020,Cao2020,bisognano_1, bisognano_2}.

Here we introduce a method for obtaining a PH, referred to as \emph{inverse quantum annealing} (IQA), which is inspired by 
quantum annealing~\cite{Kadowaki_PRE1998,Farhi_SCI01,Santoro_SCI02,Hauke_2020,aqc_review}, 
but with the role of states and Hamiltonians swapped.
Given a state $|\psi_1\rangle$, whose PH $\Ham_1$ we wish to construct, and starting from a simple state 
$|\psi_0\rangle$ with a well-known PH $\Ham_0$, we construct a path $|\psi(\lambda(t))\rangle$, connecting 
$|\psi_0\rangle$ to $|\psi_1\rangle$. We then write down an artificial dynamics for the Hamiltonian, inspired by von Neumann's equation for the density matrix, which is amenable of a well-defined adiabatic limit, and, more importantly, can be approximately solved with a space of local Hamiltonians. In the adiabatic limit we obtain a Hamiltonian having the target state as the ground state. 
We will illustrate the main ideas of the method with two paradigmatic examples: 
1) the ``exactly solvable'' Kitaev fermionic chain, where $|\psi(\lambda)\rangle$ crosses a second-order transition point, and 
2) a quantum Ising chain in the presence of a longitudinal field, where $|\psi(\lambda)\rangle$ crosses a first-order transition point.
In case 1) power-law correlations emerge, making the local approximation harder,
while in case 2) correlations are exponentially decreasing and the local approximation is excellent.  

{\em Inverse quantum annealing protocol.---} 
Given a (many-body) state $\ket{\psi_1}$, the task is to find a (local) Hamiltonian $\Ham_1$ 
for which $\ket{\psi_1}$ is the ground state. 
In many cases, this problem has more than one solution~\footnote{For instance, the standard state $|\psi_0\rangle=|+\rangle^{\otimes N}$ has a non-local parent Hamiltonian $\Hamaux=-|\psi_0\rangle 
\langle \psi_0|$, but also a quite simple 1-local PH $\Ham_0=-\sum_j\PauliSigma^x_j$.}.
To find a solution, we introduce a method inspired by Quantum Annealing~\cite{Kadowaki_PRE1998,Farhi_SCI01,Santoro_SCI02}, 
{\em alias} Adiabatic Quantum Computation~\cite{aqc_review}. 
The first step is to introduce a family of states $\ket{\psi(\lambda)}$, depending on a parameter $\lambda$, with $0\le \lambda \le 1$, such that: 
i) $\ket{\psi(1)} \equiv \ket{\psi_1}$ is the quantum state whose PH $\Ham_1$ we wish to determine, and ii) $\ket{\psi(0)} \equiv \ket{\psi_0}$ is a simple initial state whose PH $\Ham_0$ is known. 
Then we seek a dynamics that, by changing $\lambda(t)$ with time, in the adiabatic limit $\dot{\lambda}(t)\to 0$, leads to the desired $\Ham_1$, starting from $\Ham_0$. 

The idea behind this artificial Hamiltonian dynamics is the following. 
Consider the projector $\Projector_{\psi(\lambda)}$ on the selected state path $|\psi(\lambda)\rangle$, suitably redefined as follows:
\begin{equation} \label{eqn:Projector_psi}
\Projector_{\psi(\lambda)} = - J |\psi(\lambda)\rangle \langle \psi(\lambda) | \;,
\end{equation}
where $J$ is an arbitrary energy scale which we use as our unit, setting $J=1$.
$\Projector_{\psi(\lambda)}$ has $|\psi(\lambda)\rangle$ as its unique ground state, at energy $-J$, while all other states are degenerate, at energy $0$. 

Regard now $\Projector_{\psi(\lambda(t))}$ as a {\em(pseudo-)Hamiltonian} --- in general, non-local --- generating a Schr\"odinger dynamics associated to the evolution operator
\begin{equation}
\Unitary(t) = \mathrm{T}\!\!-\!\!
\exp \Big( -\textstyle{\frac{i}{\hbar} \int_0^t \! \ud t'} \; \Projector_{\psi(\lambda(t'))} \Big) \;.
\end{equation}  
By assumption a local PH $\Ham_0$ for $|\psi_0\rangle$ exists.
Consider now an ``auxiliary Hamiltonian'' $\Hamaux(t)=\Unitary(t) \Ham_0 \Unitary^\dagger(t)$.
It will satisfy von Neumann's equation:
\be \label{eq:not_projected}
\partial_t \Hamaux(t) = - \frac{i}{\hbar} \big[ \Projector_{\psi(\lambda(t))},\Hamaux(t) \big] \;,
\ee
with the boundary condition $\Hamaux(0)=\Ham_0$.
The presence of the spectral gap $J$ in $\Projector_{\psi(\lambda)}$ guarantees that, 
in the adiabatic limit $\dot{\lambda}\to 0$, the time-evolved state $\Unitary(t)|\psi_0\rangle$ will be closer and closer to the desired path of states $|\psi(\lambda(t))\rangle$, and, correspondingly, the ``Hamiltonian'' $\Hamaux(t)$ will approximate a PH $\Ham(\lambda(t))$.
The non-trivial issue with such an adiabatically-inspired solution for the PH problem is the possible {\em non-locality} 
of the Hamiltonian determined. 
We need to devise a further {\em local approximation} for the PH problem to guarantee that the solution found is actually a physical {\em local} PH.

Before tackling the locality issue, let us rewrite our equation using a fixed basis of Hermitian operators $\calP=\{\opP_{\rmj}\}$  acting on a system of $N$ particles. 
As an example, think of a system made of $N$ spin-1/2, where $\opP_{\rmj}$ are all possible Pauli string operators made by an arbitrary number of Pauli matrices $\PauliSigma_i^{x,y,z}$ 
at sites $i$. 
Without loss of generality, we can assume that the normalization of the operators is such that 
$\tr(\opP_{\rmj} \opP_{\rmj'})=\delta_{\rmj,\rmj'}$~\cite{supp}.
For any finite $N$, the total number of elements in $\calP$ is finite, $\calN=\big|\calP\big|$. 
Any arbitrary Hermitian operator can be expanded in the basis $\calP$, 
for instance, $\Hamaux(t)=\sum_{\rmj} h_j(t) \, \opP_{\rmj}$. 
By substituting in Eq.~\eqref{eq:not_projected}, after simple algebra, see~\cite{supp}, we
can rewrite 
\eqref{eq:not_projected} as
\begin{equation} \label{eqn:h_nonprojected}
\partial_t h_{\rm j}(t) = \sum_{\rmj'=1}^{\calN} K_{\rmj,\rmj'}[\psi(\lambda(t))]\, h_{\rmj'}(t) \;,
\end{equation}
where 
$K_{\rmj,\rmj'}[\psi]\equiv -i(J/\hbar) \bra{\psi}\big[\opP_{\rmj},\opP_{\rmj'}\big]\ket{\psi}$
is a skew-symmetric \textit{commutator matrix}.

The space of $l$-local Hamiltonians is formed by linear combinations of a subset 
$\calL^{(l)}\subset \calP$ consisting of all Hermitian operators connecting particles within maximum distance $l$ on a lattice, whose number we denote as $\calN_l$. (Let us call the maximum coupling distance of an operator as its coupling length.)
As an example, a $1$-local Hamiltonian contains only single-particle terms, while
a $2$-local Hamiltonian will also contain two adjacent particles interactions. In both cases, $\calN_l$ scales linearly in $N$.
Any $l$-local Hamiltonian is written as $\Ham^{(l)}=\sum_{\rmj} h_{\rmj} \opL_{\rmj}^{(l)}$.
We call a Hamiltonian {\em local} if its locality range $l$ does not depend 
on the system size $N$.

To find an optimal $l$-local PH from the adiabatic solution of Eq.~\eqref{eq:not_projected}, 
we now use a time-dependent variational principle (TDVP)~\cite{TDVP-MPS}, 
which allows us to determine the Hamiltonian couplings $h_{\rmj}(t)$, by projecting the right-hand side of Eq.~\eqref{eq:not_projected} on the space $\calL^{(l)}=\{\opL_{\rmj}^{(l)}\}$, through the Hilbert-Schmidt distance $d(\opA,\opB)=\sqrt{\tr(\opA-\opB)^2}$, a natural Euclidean structure in the space of Hermitian operators.
As detailed in~\cite{supp}, we can write the resulting projected evolution as:
\begin{equation}
\partial_t \Ham^{(l)} = \Project_l \Big( -\frac{i}{\hbar} \big[ \Projector_{\psi(\lambda(t))},\Ham^{(l)}(t) \big] \Big) \;,
\end{equation}
where the projector $\Project_l(\opA)$ defines the closest $l$-local operator $\opB$ to a given $\opA$. 
This leads to the following equation for the coefficients $h_{\rmj}(t)$ of the $l$-local Hamiltonian $\Ham^{(l)}(t)$:
\be \label{eq:main}
\partial_t h_{\rmj}(t) = \sum_{\rmj'=1}^{\calN_l} K_{\rmj,\rmj'}^{(l)} [\psi(\lambda(t))] \, h_{\rmj'}(t) \;,
\ee
where $K_{\rmj,\rmj'}^{(l)} [\psi]\equiv -i(J/\hbar) \bra{\psi}\big[\opL_{\rmj}^{(l)},\opL_{\rmj'}^{(l)}\big]\ket{\psi}$ has a size that scales polynomially with the system size $N$. Let us note that Eq.~\eqref{eq:main} is a truncated version of Eq.~\eqref{eqn:h_nonprojected}, with the commutator matrix restricted to the space of $l$-local operators.

Equation~\eqref{eq:main} is the central result of this work. 
In the adiabatic regime, it allows us to construct an $l$-local PH $\Ham_1$ 
for the final state $|\psi_1\rangle$, by integrating the differential equations from $t=0$, with initial condition set by the expansion coefficients $h_{\rmj}(0)$ of $\Ham_0$, up to a suitably large annealing time $T$.
While the non-projected adiabatic scheme behind Eqs.~\eqref{eqn:Projector_psi}-\eqref{eq:not_projected} is, by construction, protected by a gap $J$, 
there is no guarantee that the projection on the space of $l$-local Hamiltonians will not bring in components of higher states. 

The error introduced by the $l$-local approximation is inherently related to the nature of the
path of states $\ket{\psi(\lambda)}$. 
Our physical expectation is that if $\ket{\psi(\lambda)}$ has a finite correlation length, then the adiabatic solution of Eq.~(\ref{eq:not_projected}) is a local Hamiltonian. 
In this case the matrix $K_{\rmj,\rmj'}^{(l)}$ connects operators with similar coupling length, and the truncation leading from Eq.~\eqref{eqn:h_nonprojected} to Eq.~\eqref{eq:main} gives rise to a small error (akin to truncating a short-range Hamiltonian --- we discuss this point in~\cite{supp}). 
In this case, we can truncate to a coupling length $l$, independent of the system size, such that the 
error introduced by the TDVP projection is arbitrarily small. The opposite occurs if there is a $\lambda_{c}$ where the correlation length of $\ket{\psi(\lambda)}$ diverges: $K_{\rmj,\rmj'}^{(l)}$ connects operators with very different coupling lengths, and the truncation gives rise to a large error.

The commutator matrix $K_{\rmj,\rmj'}^{(l)}$ has been used in previous works to reconstruct local Hamiltonians from their eigenstates~\cite{Bairey2019}. 
Previous methods, however, could not guarantee to find Hamiltonians having $\ket{\psi(\lambda)}$ as the {\em ground state}. 
The IQA, implemented through Eq.~\eqref{eq:main}, can select a PH, without the need of an explicit diagonalization for checking the solution. 

We now illustrate our method on two paradigmatic examples: 
1) the integrable quantum Ising chain in a transverse field, where we will exploit the Jordan-Wigner mapping to 
to the 1D Kitaev chain,
illustrating the case of a diverging correlation length for $\ket{\psi(\lambda)}$, at a second-order transition;
2) the non-integrable quantum Ising chain in a transverse and longitudinal field, where the path $\ket{\psi(\lambda)}$ crosses a first-order transition, with a finite correlation length.

{\em 1. IQA with fermionic Gaussian states.---} 
To illustrate our IQA method, we apply it here to the BCS-like states
\beN
\ket{\psi(\lambda)}\equiv 
\prod_{\substack{k=(2n-1)\pi/N\\n\in\{1,\dots,N/2\}}} \left(\sin(\theta_k)+\cos(\theta_k)
\opcdag{k} \opcdag{-k}\right)\ket{0}\,,
\eeN
with
\beN
\theta_k(\lambda)=\frac{1}{2} \arctan \left( \frac{\sin\left(\lambda\pi/2\right)\sin(k)}{\cos\left(\lambda\pi/2\right)+\sin\left(\lambda\pi/2\right)\cos(k)}
\right) \,.
\eeN
Here $\ket{0}$ is the vacuum state, $\opcdag{k} = (e^{i\pi/4}/\sqrt{N}) \sum_{j=1}^N e^{ikj} \opcdag{j}$, and $\opcdag{j}$ creates a spinless fermion at  
site $j=1,\cdots N$.
$\ket{\psi(\lambda)}$ is the ground state of the 1D Kitaev model~\cite{Kitaev_2001} 
\begin{eqnarray} \Ham_{\text{K}}(\lambda) &=&
\sum_{j=1}^{N}\Big[ \sin\big( \lambda\frac{\pi}{2} \big)
\big( \opcdag{j} \opcdag{j+1} + \opcdag{j} \opc{j+1} + \text{h.\,c.}\big) \nonumber \\
&& \hspace{5mm} + \cos\big( \lambda\frac{\pi}{2} \big) 
\big( \opcdag{j} \opc{j} - \opc{j} \opcdag{j} \big)
\Big] \;, 
\end{eqnarray}
with anti-periodic boundary conditions $\opc{N+1}=-\opc{1}$, hence a $2$-local PH exists.
The goal is to use the dynamics defined in Eq.~\eqref{eq:main} to find a PH,
using the exactly known results to quantify the accuracy of the IQA.
The annealing schedule is $\lambda(t)=t/T$, where $T$ is the final time so that the state interpolates between $\ket{\psi(0)}$, the ground state of $\Ham_\text{K}(0)$, 
and $\ket{\psi(1)}$, the ground state of $\Ham_\text{K}(1)$.  
Note that $\ket{\psi(\lambda)}$ passes through an emerging second-order phase transition critical point at $\lambda_c=1/2$ where the correlation length diverges. 
Therefore, as a consequence of the TDVP approximation, we expect IQA to work very well for paths such that $\lambda<\lambda_c$  while it may be less accurate if $\lambda \ge \lambda_c$.

We perform IQA with different annealing times $T$ to study the convergence to the adiabatic limit, analyzing different ranges 
$l$ of the interactions in $\mathcal{L}^{(l)}$, and different system sizes $N$. The basis $\mathcal{L}^{(l)}$ of translation and reflection invariant $l$-interacting quadratic fermions is
$\{\Sigma_0^z/\sqrt{2},\Sigma_1^{\alpha},\dots,\Sigma_{l}^{\alpha'},\dots\}$ for $l<N/2$,
and
$\mathcal{L}^{(N/2-1)}\cup\{\Sigma_{N/2}^x/\sqrt{2},\Sigma_{N/2}^y/\sqrt{2}\}
$ for $l=N/2$,
where $\alpha, \alpha' \in\{ x,y,z\}$ and
$\Sigma_m^x= (1/2\sqrt{N})\sum_{j} 
(\opcdag{j} \opcdag{j+m} +\text{h.\,c.})$, 
$\Sigma_m^y= (i/2\sqrt{N})\sum_{j} 
(\opcdag{j} \opcdag{j+m} - \text{h.\,c.})$, 
and 
$\Sigma_m^z=(1/2\sqrt{N})\sum_{j} 
(\opcdag{j} \opc{j+m} + \text{h.\,c.})$. 
The antiperiodic boundary conditions imply $\opc{N+m}\equiv -\opc{m}$. The commutator matrix that generates the IQA is explicitly calculated in~\cite{supp}, where we also investigate the locality of the adiabatic solution of Eq.~\eqref{eq:not_projected}.

\begin{figure}[t]
    \includegraphics[height=5.6cm]{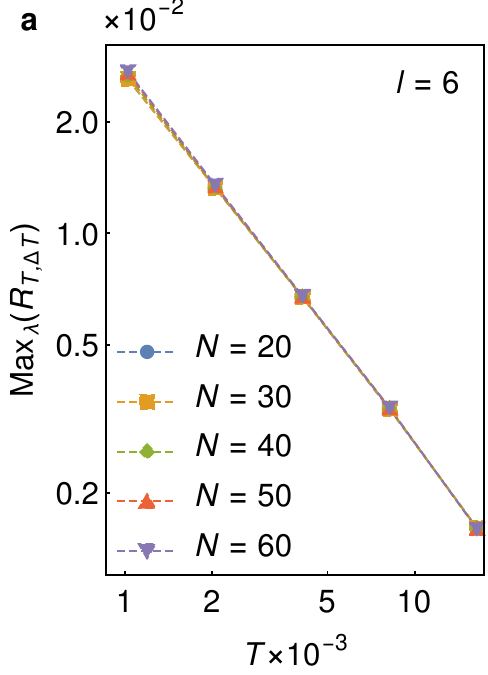} 
    \hspace{2mm}
    \includegraphics[height=5.6cm]{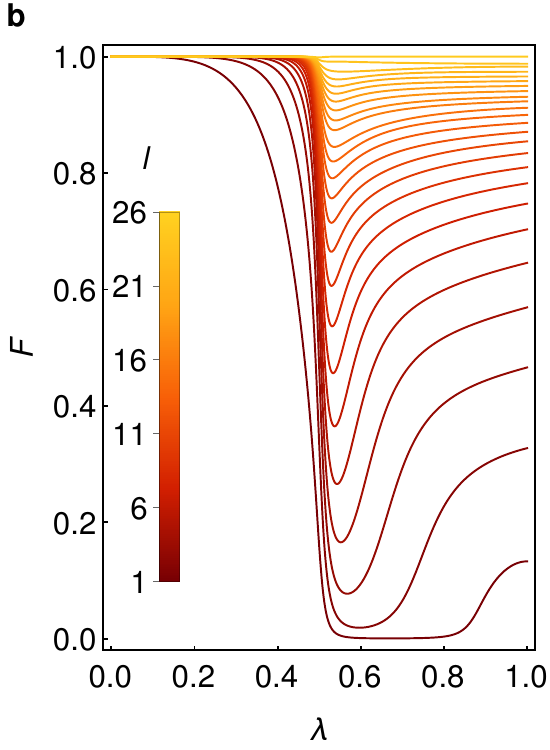}
\caption{IQA with fermionic Gaussian states --- In Panel (a): Maximum value $\text{Max}_\lambda\left(R_{T,\Delta T}(\lambda)\right)$ of the relative distance $R_{T,\Delta T}(\lambda)$ between solutions of Eq.~\eqref{eq:main} with different annealing times $T$ and $T+\Delta T = 2T$, for $l=6$, as a function of the annealing time, for systems of different sizes. In Panel (b): Fidelity between the target state $\ket{\psi(\lambda)}$ and the ground state $\ket {\psi_\text{GS}^{(l)}(\lambda)}$ of the adiabatic $l$-local Hamiltonian $\Ham^{(l)}(\lambda)$ obtained via IQA. 
We consider a system of $50$ sites and interaction ranges from $l=1$ to $l=26$.}
    \label{fig:adiabatic_time_and_fidelity}
\end{figure}

\begin{figure*}[t]
    \includegraphics[height=5.1cm]{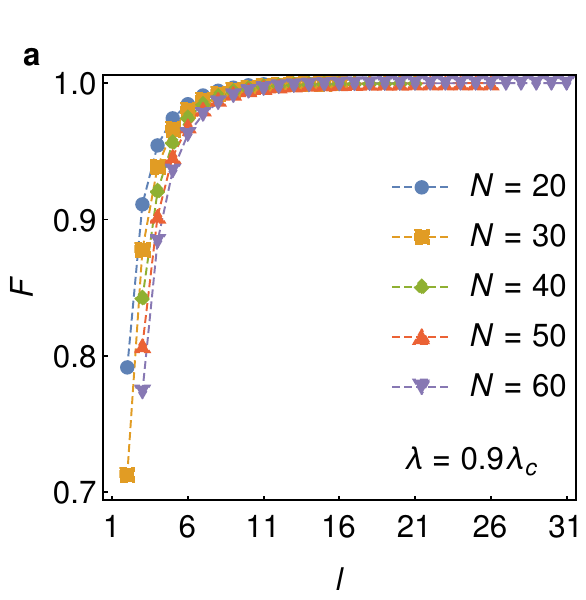} 
    \hspace{0mm}
    \includegraphics[height=5.1cm]{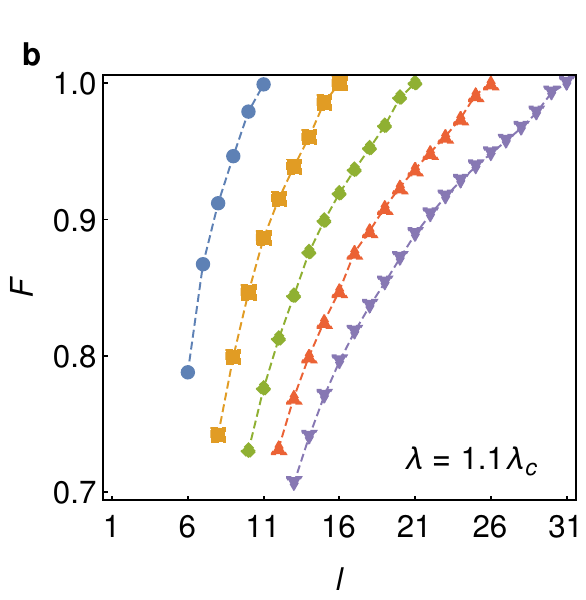}
    \hspace{0mm}
    \includegraphics[height=5.1cm]{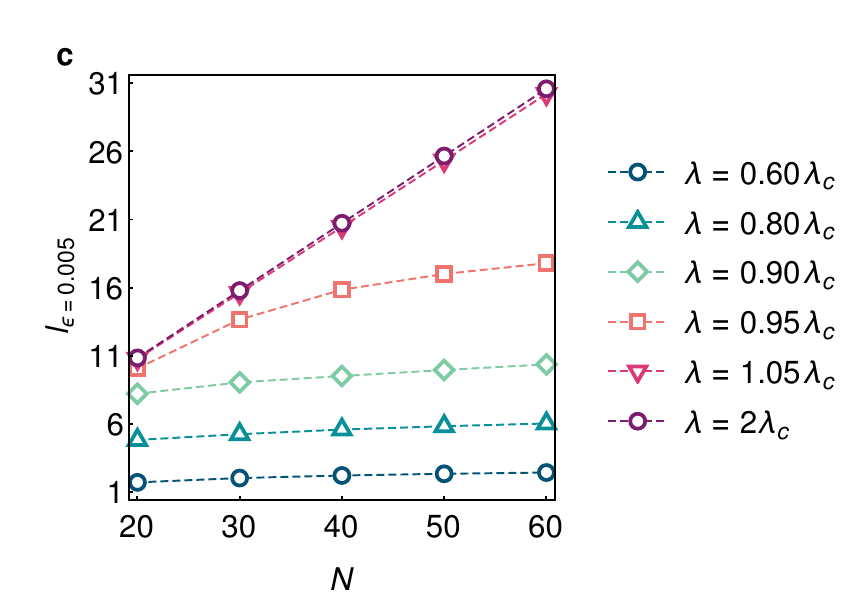}
\caption{IQA with fermionic Gaussian states --- In Panels (a) and (b): fidelity between the target state $\ket{\psi(\lambda)}$ and the ground state $\ket {\psi_\text{GS}^{(l)}(\lambda)}$ of the adiabatic $l$-local Hamiltonian $\Ham^{(l)}(\lambda)$ as a function of the interaction range $l$ and for different system sizes $N$ (legend in Panel(a)). In Panel (a) $\lambda$ is just before the phase transition, i.e. at $\lambda=0.9\lambda_c$, in Panel (b) just after the phase transition, i.e. at $\lambda=1.1\lambda_c$. In Panel (c): Minimal interaction range $l_\epsilon$ required to ensure a fidelity $F\geq 1-\epsilon$ ($\epsilon = 0.005$) between the target state and the ground state of the Hamiltonian obtained via the IQA, as a function of the system size and for different values of $\lambda$.}
    \label{fig:FvsL_and_l_espilon}
\end{figure*}

\begin{figure}
    \includegraphics[height=5.3cm]{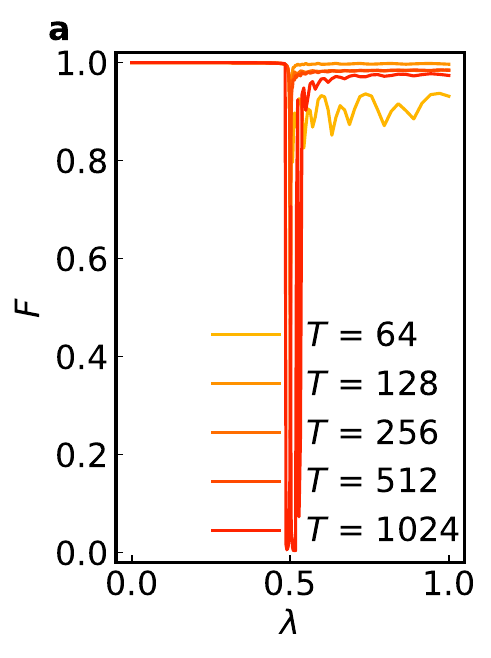} 
    \hspace{0mm}\includegraphics[height=5.3cm]{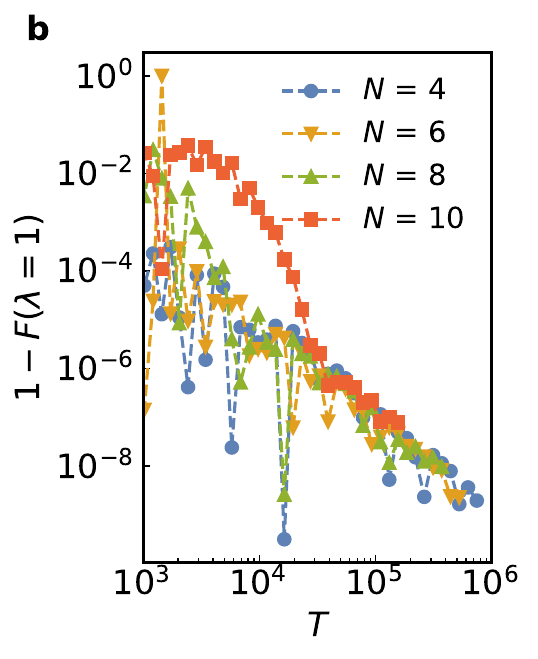}
\caption{IQA with a non-integrable Ising chain  ($B_z=0.8$) --- In Panel (a), fidelity between the target state $\ket{\psi(\lambda)}$ and the ground state $\ket {\psi_\text{GS}^{(l)}(\lambda)}$ of the adiabatic $2$-local Hamiltonian, for a system of 10 spins and different annealing times. In Panel (b), final infidelity at $\lambda=1$ as a function of the annealing time, for different system sizes.}
    \label{fig:non-integrable-0.8}
\end{figure}

To verify when the adiabatic evolution is achieved, we compute the relative Hilbert-Schmidt operator distance
\begin{equation} \label{eqn:R}
	R_{T,\Delta T}(\lambda)=\sqrt{\tr\left( \Ham_{T+\Delta T}(\lambda)-\Ham_{T}(\lambda)\right)^2/\tr\Ham^2_{T}(\lambda)}
\end{equation}
between the Hamiltonians $\Ham_{T}(\lambda)$ found by IQA, at fixed $l$ and different final times $T$ and $T+\Delta T$. Since the operators in $\mathcal{L}$ are orthonormal, this is the distance between the couplings vectors $(h_1,h_2,\dots)$ of the two Hamiltonians. When the annealing time $T$ is sufficiently large, the couplings $\{h_{\rmj}\}$ converge to those of the adiabatic Hamiltonian and $R_{T,\Delta T}(\lambda)$ goes to zero. In Fig.~\ref{fig:adiabatic_time_and_fidelity} (a), we show the maximum 
value of $R_{T,\Delta T}(\lambda)$, for different annealing times and system sizes. The functions $\text{Max}_\lambda(R_{T,\Delta T}(\lambda))$ for different values of $N$ overlap and 
fit to $\text{Max}_\lambda(R_{T,\Delta T}(\lambda))\propto 1/T$. The error is inversely proportional to the annealing time and is independent of the system size because the expectation values corresponding to the entries of the matrix $K_{\rmj,\rmj'}^{(l)}$  converge for large $N$. Having determined the adiabatic regime, we fix $T=16000$ and investigate the properties of the PH obtained from the IQA.

In Fig.~\ref{fig:adiabatic_time_and_fidelity} (b), we plot the fidelity $F(\lambda)=\lvert\langle\psi_\text{GS}^{(l)}(\lambda)\ket{\psi(\lambda)}\rvert^2$ between the target state $\ket{\psi(\lambda)}$ and 
the unique ground state $\ket {\psi_\text{GS}^{(l)}(\lambda)}$ of the adiabatic $l$-local Hamiltonian $\Ham^{(l)}(\lambda)$ obtained from the IQA.  If  $\lambda < \lambda_c=1/2$ the fidelity $F(\lambda)$ is close to $1$ 
even for relatively small values of $l$. This means that our algorithm finds an optimal $l$-local PH for the target state. 
The scenario changes when $\lambda>\lambda_c$. Indeed, as we show in~\cite{supp}, the capability of the evolution in Eq.~\eqref{eqn:h_nonprojected} of connecting local and non-local operators, as well as the effective interaction range of its adiabatic solution, scales with the correlation length. As a consequence, the effects of long-range correlations can be accounted by the TDVP only by increasing the value of $l$. This happens even though the Kitaev Hamiltonian is $2$-local, and is reminiscent of what happens in quantum annealing at phase transitions~\cite{PhysRevA.69.062302}.

A more quantitative analysis is obtained by looking at the fidelity as a function of $l$ for different system sizes. If we restrict our target state to $\lambda<\lambda_c$, the fidelity is close to $1$ 
even at small values of $l$, and almost independent of $N$, see Fig.~\ref{fig:FvsL_and_l_espilon} (a). In Fig.~\ref{fig:FvsL_and_l_espilon} (b), the fidelity is shown for a target state $\ket{\psi(\lambda)}$ with $\lambda$ close but beyond $\lambda_c$, $\lambda=1.1\lambda_c$. In this case, the larger $l$ the better the fidelity, as expected.

In order to better quantify when $\Ham^{(l)}(\lambda)$ is a suitable PH, let us fix a target accuracy $\epsilon$. We can accept $\Ham^{(l)}(\lambda)$ as a PH if the fidelity between $\ket{\psi(\lambda)}$ and the ground state $\ket{\psi_\text{GS}^{(l)}(\lambda)}$ of $\Ham^{(l)}(\lambda)$, $F_l(\lambda)=|\langle{\psi(\lambda)}\ket{\psi_\text{GS}^{(l)}(\lambda)}|^2 $ is larger than $1-\epsilon$.
This condition defines a minimal interaction range $l_\epsilon$ required to adiabatically find the PH within the given accuracy $\epsilon$, i.\,e., $F_l(\lambda) \geq1-\epsilon$ $\forall l\geq l_\epsilon$.
This length $l_\epsilon$ is plotted versus $N$ in Fig.~\ref{fig:FvsL_and_l_espilon} (c) for $\epsilon=0.005$ and different values of $\lambda$. For $\lambda<\lambda_c$, $l_\epsilon$ weakly depends on the system size, while, for $\lambda>\lambda_c$, $l_\epsilon$ scales almost linearly with the system size.

{\em 2. IQA with a non-integrable Ising chain.---} We apply the IQA on a path of ground states $\ket{\psi(\lambda)}$ of a $1-$dimensional Ising chain in transverse and longitudinal field 
\begin{equation}
\Ham_\text{I}(\lambda) = \sum_{j=1}^N\left(\PauliSigma_j^x\PauliSigma_{j+1}^x - B_z \PauliSigma_j^z + B_x(\lambda) \PauliSigma_j^x\right) \;,
\end{equation} 
where $\PauliSigma^{x,z}_j$ are Pauli matrices, and $\PauliSigma_{N+1}^x\equiv\PauliSigma_{1}^x$. 
We choose $B_x(\lambda)=0.9\cdot(\lambda-\lambda_c)$ with $\lambda_c=0.5$. Here, a first-order quantum phase emerges at $\lambda=\lambda_c$ and allows us to further challenge our method. We perform the IQA for $\lambda \in [0,1]$, $N\in\{4,6,8,10\}$ and $\mathcal{L}^{(2)}$ containing all the local translationally invariant operators up to range $l=2$. The time-schedule is $\lambda(t) = 4(t/T)^3 - 6(t/T)^2 + 3(t/T)$ and has null derivative when $\lambda=\lambda_c$ to slowly cross the emerging criticality.

Results of the IQA are summarized in Fig.~\ref{fig:non-integrable-0.8} for $B_z=0.8$ (see~\cite{supp} for $B_z=0.9$) corroborating our considerations regarding the effectiveness of our method and the effect of different order phase transitions on the adiabatic time and on the error introduced by the TDVP. 
In Panel (a) we plot the fidelity $F(\lambda)$ between the target state and the ground state of the IQA Hamiltonian, as a function of the control parameter $\lambda$ for different annealing times. We see that before the phase transition at $\lambda_c$, a large fidelity is reached also for short annealing times. 
Near $\lambda_c$ the fidelity drops, but, differently from the second order transition case, we observe a large recovery after $\lambda_c$.

In Fig.~\ref{fig:non-integrable-0.8} (b), where we plot the final infidelity $1-F(\lambda)$ at $\lambda = 1$ for different annealing times. We observe that after the critical point, 
the infidelity decreases polynomially with the annealing time and becomes independent on the system size. Remarkably, a small interaction range $l=2$ is sufficient to construct an excellent PH in the adiabatic limit.

{\em Discussion and conclusions ---} Quantum annealing represents one of the major examples of the computational potential of quantum many-body systems. In this work, exploiting a combination of adiabatic approximation and TDVP, we introduced an annealing technique for approximating the $l$-local parent Hamiltonian of a target many-body state. We exemplified our method by reconstructing PHs for the Kitaev fermionic chain and a quantum Ising chain in longitudinal and transverse fields. The IQA allows for efficient reconstruction of PHs for paths of many-body states with finite correlation lengths, independently of the integrability of the model or the emergence of a first-order phase transition. The next step is to use IQA to design Hamiltonians that would allow for the experimental preparation of quantum states relevant in many-body physics and quantum information, directly from their wavefunctions. Remarkable examples include quantum spin liquid variational wavefunctions~\cite{Zhou_2017}.
Finally, IQA only relies on the knowledge of local expectation values. This can be relevant for applications to the quantum marginal problem~\cite{Xin2019,PhysRevLett.89.277906,q_marginal_p_1,q_marginal_p_2,marginal_0,marginal_1,marginal_2}.

\begin{acknowledgments}
We acknowledge M. Dalmonte for his very useful suggestions and comments on the manuscript. 
G.~P. and P.~L. acknowledge financial support from the project PIR01\_00011 “IBiSCo”, PON 2014-2020.
G.~E.~S., R.~F., P.~L. and A.~R. acknowledge financial support from PNRR MUR project PE0000023-NQSTI. 
G.~E~.S. and P.~L. acknowledge financial support from the project QuantERA II Programme STAQS project that has received funding from the European Union’s Horizon 2020 research and innovation programme under Grant Agreement No 101017733. A.~R. thanks SISSA for the warm hospitality received during the completion of this work.
This work is co-funded by the European Union (ERC, RAVE, 101053159). 
Views and opinions expressed are however those of the author(s) only and do not necessarily reflect those of the European Union or the European Research Council. Neither the European Union nor the granting authority can be held responsible for them.
\end{acknowledgments}

\bibliography{refs}

\clearpage

\onecolumngrid
\appendix

\setcounter{equation}{0}
\setcounter{figure}{0}
\setcounter{table}{0}
\setcounter{page}{1}
\renewcommand{\theequation}{S\arabic{equation}}
\renewcommand{\thefigure}{S\arabic{figure}}
\renewcommand{\bibnumfmt}[1]{[S#1]}
\renewcommand{\citenumfont}[1]{#1}

\begin{adjustwidth}{20pt}{20pt}
\section{ Parent Hamiltonian reconstruction via inverse quantum annealing - Supplementary Material}

\subsection{The $l$-local projection and the time-dependent variational principle}

Let us consider basis of Hermitian operators $\calP=\{\opP_{\rmj}\}$ acting on a system of $N$ particles. We choose the elements in $\calP$ end their normalization such that
\be \label{eqn:P_orthonormality}
\tr[ \opP_{\rmj} \opP_{\rmj'} ]=\delta_{\rmj,\rmj'} \;,
\ee 
As an example, for a system made of $N$ spin-1/2, we can define the $\opP_{\rmj}$ as the set of all possible Pauli string operators $\opP = \frac{1}{\sqrt{2^N}} \, \PauliSigma^{\alpha_1}_1 \, \PauliSigma^{\alpha_2}_2 \, \cdots \, \PauliSigma^{\alpha_n}_n $ where $\alpha_j=x,y,z$ for the ordinary Pauli matrices $\PauliSigma_j^{x,y,z}$ at site $j$, while $\alpha_j=0$ denotes the identity, $\PauliSigma_j^{0}=\identity_j$.

A $N$-particles operator on a lattice is termed \emph{$l$-local} if it acts differently that the identity only on particles within a maximum lattice distance $l$. For example, the first-neighbouring interaction $\opP = \frac{1}{\sqrt{2^N}} \, \PauliSigma^{x}_1 \, \PauliSigma^{y}_2 \, \identity_3 \, \cdots \, \identity_3 $ on a spin lattice is a $2$-local operator.

Any Hermitian operator $\opA$ can be expanded in terms of elements of $\calP$.
Given a value of $l<n$, you could ask what is the $l$-local operator $\opB \in \textrm{span}(\calL^{(l)})$, 
that is closest to $\opA$ according to some distance $d(\opA,\opB)$.
We consider here the trace distance $d(\opA,\opB)=\sqrt{ \tr[(\opA-\opB)^2]}$.
By conveniently organizing the indices, the expansion of $\opA$ in terms of a set of real coefficients 
$a_{\rmj}$ can be equivalently expressed as:
\be \label{eqn:A_expansion}
\opA = \sum_{\rmj=1}^{\calN} a_{\rmj} \, \opP_{\rmj} 
= \sum_{\rmj=1}^{\calN_{l}} a_{\rmj} \, \opL^{(l)}_{\rmj}  + \sum_{\rmj>\calN_{l}}^{\calN} 
a_{\rmj} \, \opP_{\rmj} \;.
\ee
By expressing the desired $l$-local approximation $\opB$ as
\[
\opB = \sum_{\rmj=1}^{\calN_{l}} b_{\rmj} \, \opL_{\rmj}^{(l)} \;,
\]
and imposing a vanishing gradient for $\tr[(\opA-\opB)^2]$, we deduce that:
\begin{eqnarray}
0 = \frac{\partial}{\partial b_{\rmj}} \tr[(\opA-\opB)^2] 
&=& \frac{\partial}{\partial b_{\rmj}} 
   \Big( \tr\big[\opA^2\big] -2\sum_{\rmj=1}^{\calN_{l}} b_{\rmj} \tr\big[ \opA \, \opL^{(l)}_{\rmj}\big] + 
   \sum_{\rmj=1}^{\calN_{l}} \sum_{\rmj'=1}^{\calN_{l}} 
   b_{\rmj} b_{\rmj'} \tr\big[ \opL^{(l)}_{\rmj} \opL^{(l)}_{\rmj'} \big]
   \Big) \nonumber \\
&=& 2 \Big( \sum_{\rmj'=1}^{\calN_{l}} b_{\rmj'} \tr\big[ \opL^{(l)}_{\rmj} \opL^{(l)}_{\rmj'} \big] - 
\tr\big[ \opA \, \opL^{(l)}_{\rmj}\big] \Big) \nonumber \\
&=& 2 \Big( b_{\rmj} - \tr\big[ \opA \, \opL^{(l)}_{\rmj}\big] \Big) 
\hspace{10mm} \Longleftrightarrow \hspace{10mm} b_{\rmj} = \tr\big[ \opA \, \opL^{(l)}_{\rmj}\big] 
\equiv a_{\rmj} \;,
\end{eqnarray}
where we used the orthonormality condition \eqref{eqn:P_orthonormality} for the operators in $\calL^{(l)}$, 
and the expansion of $\opA$ in Eq.~\eqref{eqn:A_expansion}. We deduced a simple result: in words, the $l$-local operator $\opB$ which is closest in trace-distance to a given operator $\opA$ is simply obtained by setting to zero all the coefficients $a_{\rmj}$ of Pauli string operators 
$\opP_{\rmj}$ not belonging to the desired $l$-local set $\calL^{(l)}$, 
while $b_{\rmj}=a_{\rmj}$ for the $l$-local terms. 
This is essentially, a {\em projection} on the space of $l$-local operators spanned by $\calL^{(l)}$, which we
denote as
\be
\Project_{l} (\opA) =  \sum_{\rmj=1}^{\calN_{l}} a_{\rmj} \, \opL^{(l)}_{\rmj}  \;.
\ee

Now, take the ``Hamiltonian'' $\Hamaux(t)$
considered in the main text.
We can project it on the space of $l$-local operators $\calL^{(l)}$:
\be
\Project_{l} \big(\Hamaux(t) \big) =  \sum_{\rmj=1}^{\calN_{l}} h_{\rmj}(t) \, \opL^{(l)}_{\rmj} 
\defuguale \Ham^{(l)}(t) \;,
\ee
with appropriate real coefficients $h_{\rmj}(t)$. 
The dynamics imposed by the von Neumann's equation associated to the projector $\Projector_{\psi}$ 
on the state $|\psi(\lambda)\rangle$, however, brings the $l$-local projected $\Ham^{(l)}(t)$ {\em out} of the 
desired space spanned by $\calL^{(l)}$. 
In a way that is reminiscent of the time-dependent variational principle (TDVP) idea~\cite{TDVP-MPS}, 
we can project again the result on the $l$-local space, obtaining an $l$-local projected dynamics of the
form:
\be \label{eqn:projected}
\partial_t \Ham^{(l)}(t) = - \frac{i}{\hbar} \Project_{l} \Big( \big[ \Projector_{\psi(\lambda(t))},\Ham^{(l)}(t) \big] \Big) \;.
\ee
To determine the differential equations for the coefficients $h_{\rmj}(t)$, let us project the commutator
\[
-\frac{i}{\hbar} \big[ \Projector_{\psi(\lambda(t))},\Ham^{(l)}(t) \big] = 
-\frac{i}{\hbar} \sum_{\rmj'=1}^{\calN_{l}} h_{\rmj'}(t) \, \big[ \Projector_{\psi(\lambda(t))},\opL^{(l)}_{\rmj'} \big]  
\] 
on the subspace spanned by $\calL^{(l)}$, by multiplying by $\opL^{(l)}_{\rmj}$ and taking the trace:
\begin{eqnarray}
-\frac{i}{\hbar} \sum_{\rmj'=1}^{\calN_{l}} h_{\rmj'}(t) \, \tr \Big( \opL^{(l)}_{\rmj} \big[ \Projector_{\psi(\lambda(t))},\opL^{(l)}_{\rmj'} \big]  \Big) &=& 
\frac{i}{\hbar} \sum_{\rmj'=1}^{\calN_{l}}  \tr \Big( \big[ \opL^{(l)}_{\rmj}, \opL^{(l)}_{\rmj'} \big]  \Projector_{\psi(\lambda(t))}  \Big) \, h_{\rmj'}(t) \nonumber \\
&=& -\frac{iJ}{\hbar} \sum_{\rmj'=1}^{\calN_{l}}  \Big( \langle \psi(\lambda(t)) | \big[ \opL^{(l)}_{\rmj}, \opL^{(l)}_{\rmj'} \big]  |\psi(\lambda(t))\rangle \Big) \, h_{\rmj'}(t) \;.
\end{eqnarray}
Summarizing, Eq.~\eqref{eqn:projected} translates into the following differential equations for the local Hamiltonian coefficients $h_{\rmj}(t)$:
\be \label{eqn:h_projected}
\partial_t h_{\rmj}(t) =   \sum_{\rmj'=1}^{\calN_{l}} K_{\rmj,\rmj'}^{(l)}[\psi(\lambda(t))] \,
h_{\rmj'}(t) \;,
\ee
where the truncated $\calN_{l}\times \calN_l$ commutator matrix is given by:
\be \label{eqn:K_projected}
K_{\rmj,\rmj'}^{(l)}[\psi] =   
-\frac{iJ}{\hbar} \langle \psi | \big[ \opL^{(l)}_{\rmj}, \opL^{(l)}_{\rmj'} \big]  |\psi\rangle \;.
\ee 
When no $l$-local projection is performed, i.e., for $l=n$, then Eq.~\eqref{eqn:h_projected} reduces 
to Eq.~(4) of the main text, and Eq.~\eqref{eqn:K_projected} to the associated full commutator matrix.

\subsection{IQA with fermionic Gaussian states: basis of quadratic fermionic Hamiltonians}\label{app:basis}

We define the basis $\mathcal{L}^{(l)}$ of a space of translation and reflection invariant interactions of range $l$ as follows:
\begin{align*}
\mathcal{L}^{(l<N/2)}\equiv\{\Sigma_0^z/\sqrt{2},\Sigma_1^x,\Sigma_1^y,\Sigma_1^z,\dots,\Sigma_{l}^x,\Sigma_{l}^y,\Sigma_{l}^z\},
\end{align*}
and
\begin{align*}
\mathcal{L}^{(N/2)}\equiv\mathcal{L}^{(N/2-1)}\cup\{\Sigma_{N/2}^x/\sqrt{2},\Sigma_{N/2}^y/\sqrt{2}\},
\end{align*}
where
\begin{align}
\Sigma_m^x&=\frac{1}{2\sqrt{N}}\sum_{j=1}^{N} \left( \opcdag{j}  \opcdag{j+m} -  \opc{j}  \opc{j+m} \right)\nonumber\\
\Sigma_m^y&=\frac{i}{2\sqrt{N}}\sum_{j=1}^{N}\left( \opcdag{j} \opcdag{j+m}+ \opc{j} \opc{j+m}\right)\nonumber\\
\Sigma_m^z&=\frac{1}{2\sqrt{N}}\sum_{j=1}^{N} \left(\opcdag{j} \opc{j+m} -  \opc{j} \opcdag{j+m}\right)
\end{align}
with $\opcdag{j}$ and $\opc{j}$ fermionic creation and annihilation operators at position $j$ on the lattice, and antiperiodic boundary conditions $\opc{N+m}\equiv -\opc{m}$.

The operators in $\mathcal{L}^{(l)}$ allow for a common block-diagonal representation~\cite{IsingSantoro}. As a first step in this direction, we introduce the fermionic operators $\opc{k}$ in reciprocal space:
\begin{align*}
	 \opc{k} &=\frac{e^{-i\pi/4}}{\sqrt{N}}\sum_{j=1}^Ne^{-ikj} \opc{j} \;.
\end{align*}
The antiperiodic boundary conditions $\opc{N+m}=-\opc{m}$ imply that $\sum_{k\in\mathcal{K}}e^{ik(N+m)} \opc{k} =-\sum_{k\in\mathcal{K}}e^{ikm} \opc{k}$, hence
\beN
k\in \mathcal{K}\equiv\Big\{k=\pm \frac{(2n-1)\pi}{N},\text{ with }n\in\{1,\dots,\textstyle{\frac{N}{2}} \}\Big\}.
\eeN
Now the operators in $\mathcal{L}^{(l)}$ can be written as:
\begin{align}
\Sigma_{m}^x&=-\frac{i}{2\sqrt{N}}\sum_{k\in\mathcal{K}}
(e^{imk} \opcdag{k} \opcdag{-k} - e^{-imk}  \opc{-k}   \opc{k})
=\frac{1}{\sqrt{N}}\sum_{k\in\mathcal{K}} \sin(mk)
(\opcdag{k} \opcdag{-k} + \opc{-k} \opc{k}) \nonumber\\
\Sigma_{m}^y&=\frac{1}{2\sqrt{N}}\sum_{k\in\mathcal{K}} (e^{imk} \opcdag{k}   \opcdag{-k} + e^{-imk}  \opc{-k} \opc{k})
=\frac{i}{\sqrt{N}} \sum_{k\in\mathcal{K}} \sin(mk) 
( \opcdag{k} \opcdag{-k} - \opc{-k} \opc{k}) \nonumber\\
\Sigma_{m}^z&=\frac{1}{2\sqrt{N}}\sum_{k\in\mathcal{K}} e^{imk} 
( \opcdag{k} \opc{k} -  \opc{-k} \opcdag{-k}) = 
\frac{1}{\sqrt{N}}\sum_{k\in\mathcal{K}} \cos(mk) 
(\opcdag{k} \opc{k} - \opc{-k} \opcdag{-k} ) \;,
\end{align}
where we have used the canonical anti-commutation relations.
\ignore{
\begin{align*}
 \opcdag{k}  c_{k}+  \opcdag{-k}  c_{-k} &=  \opcdag{-k}  c_{-k} +   \opcdag{k}  c_{k}\\
 \opcdag{k}   \opcdag{-k} &= -   \opcdag{-k}  \opcdag{k}\\
c_k  c_{-k} &= -  c_{-k} c_k \;.
\end{align*}
}

Finally, let us define the pseudospin operators 
$\tilde\sigma_k^x = (\opcdag{k} \opcdag{-k} + \opc{-k} \opc{k})$, $\tilde\sigma_k^y=-i (\opcdag{k} \opcdag{-k} - \opc{-k} \opc{k})$, and 
$\tilde\sigma_k^z=(\opcdag{k} \opc{k} - \opc{-k} \opcdag{-k})$. These operators form, within the subspace of empty and paired fermionic states, a representation of the spin algebra on the lattice of momenta, i.\,e., 
$\{\tilde\sigma_k^\mu,\tilde\sigma_l^\nu\} =2\id\delta_{kl}\delta_{\mu\nu}$ 
and $[\tilde\sigma_k^\mu,\tilde\sigma_l^\nu]
=2i\varepsilon_{\mu\nu\gamma}\delta_{kl}\tilde\sigma_k^\gamma$, where $\varepsilon_{\mu\nu\gamma}$ is the Levi-Civita symbol. 
In terms of pseudospins, the operators in $\mathcal{L}^{(l)}$ read
\begin{align}\label{eq:psuedospins}
	\Sigma_m^x&=\frac{2}{\sqrt{N}}\sum_{k\in\mathcal{K}^+} \sin(mk)\, \tilde\sigma_k^x\nonumber\\
    \Sigma_m^y&=\frac{-2}{\sqrt{N}}\sum_{k\in\mathcal{K}^+} \sin(mk)\, \tilde\sigma_k^y\nonumber\\
    \Sigma_m^z&=\frac{2}{\sqrt{N}}\sum_{k\in\mathcal{K}^+} \cos(mk) \, \tilde\sigma_k^z,
\end{align}
where $\mathcal{K}^+\equiv\Big\{k=(2n-1)\pi/N,\text{ with }n\in\{1,\dots,N/2\}\Big\}$.

Now, we show that the operators in $\mathcal{L}^{(N/2)}$ are orthogonal and that they have the same norm.
First, let us remark that $\tr(\tilde\sigma_k^\mu\tilde\sigma_{k'}^\nu)=\tr(\{\tilde\sigma_k^\mu,\tilde\sigma_{k'}^\nu\})/2=\tr(\id)\delta_{kk'}\delta_{\mu\nu}$. It follows that the scalar products of the operators $\Sigma_m^\mu$ are
\be
\tr\left(\Sigma_m^\mu\Sigma_l^\nu\right)=\frac{4}{N}\sum_k\sum_{k'}\tr(\tilde\sigma_k^\mu\tilde\sigma_{k'}^\nu)f^\mu(km)f^\nu(k'l)=\frac{4}{N}\delta_{\mu\nu}\tr(\id)\sum_kf^\mu(km)f^\mu(kl),
\ee
where $f^\mu(x)\equiv(\sin(x),-\sin(x),\cos(x))$. Considering the orthogonality of the Fourier basis, the latter equation becomes
\begin{align}
\tr\left(\Sigma_m^\mu\Sigma_l^\nu\right)&=\delta_{\mu\nu}\delta_{ml}\frac{1}{2}\tr(\id)\qquad\text{for}\quad 0<m<N/2\nonumber\\
\tr\left(\Sigma_m^\mu\Sigma_l^\nu\right)&=\delta_{\mu\nu}\delta_{ml}\tr(\id)\qquad\text{for}\quad m\in\{0,N/2\} \;.
\end{align}
Therefore,
\be
\mathcal{L}^{(N/2)}\equiv\{\Sigma_0^z/\sqrt{2},\Sigma_1^x,\Sigma_1^y,\Sigma_1^z,\dots,\Sigma_{N/2-1}^x,\Sigma_{N/2-1}^y,\Sigma_{N/2-1}^z,\Sigma_{N/2}^x/\sqrt{2},\Sigma_{N/2}^y/\sqrt{2}\},
\ee
is a set of $3N/2$ orthogonal vectors with the same norm.

\subsection{IQA with fermionic Gaussian states: Hamiltonian diagonalization and commutator matrix}\label{app:comm_matrix}

Here we calculate the commutator matrix 
$K^{(l)}_{\rmj,\rmj'}(\lambda)$, i.e., more explicitly
$K^{(l)}_{\{n,\nu\}\{m,\mu\}}(\lambda)\equiv\bra{\psi(\lambda)}-i[\Sigma_m^\mu,\Sigma_{n}^{\nu}]\ket{\psi(\lambda)}$. These are the commutators expectation values of the operators in $\mathcal{L}^{(l)}$ for the ground states $\ket{\psi(\lambda)}$ of
\be
\Ham_{\text{K}}(\lambda) =
\sin\big( \lambda\frac{\pi}{2} \big)
\sum_{j=1}^{N} 
\big( \opcdag{j} \opcdag{j+1} + \opcdag{j} \opc{j+1} + \text{h.\,c.}\big) 
+ \cos\big( \lambda\frac{\pi}{2} \big) 
\sum_{j=1}^{N}
\big( \opcdag{j} \opc{j} - \opc{j} \opcdag{j} \big) \;, 
%
\ee
Exploiting Eqs.~\eqref{eq:psuedospins}, this Hamiltonian can be written in pseudospin form as
\be
\Ham_{\text{K}}(\lambda) =-\sum_{k\in\mathcal{K}^+}\epsilon_k
\left( v_k^x \tilde\sigma_k^x + v_k^z \tilde\sigma_k^z \right),
\ee
where
\begin{align}
\epsilon_k&=\sqrt{1+ 2\sin\left(\lambda\pi/2\right)\cos\left(\lambda\pi/2\right)\cos(k)}\nonumber\\
v_k^x&=-\sin\left(\lambda\pi/2\right)\sin(k)/\epsilon_k\\\nonumber
v_k^z&=- \left(\cos\left(\lambda\pi/2\right) + \sin\left(\lambda\pi/2\right)\cos(k)\right)/\epsilon_k \;.
\end{align}

The density matrix of the ground states of $H^\text{K}(\lambda)$ is
\be\label{eq:gss}
\rho(\lambda)\equiv\bigotimes_{k\in\mathcal{K}^+} \frac{1}{2} \left(v_k^x\tilde\sigma_k^x+v_k^z\tilde\sigma_k^z+\id\right) \,.
\ee

We will exploit this density matrix representation of the state to evaluate the commutator matrix. The corresponding state $\ket{\psi(\lambda)}$ is the $+1$ eigenvector of $\rho(\lambda)$: 
\be
\ket{\psi(\lambda)}\equiv\prod_{k\in\mathcal{K}^+} \left(\cos(\theta_k(\lambda))\, \tilde\sigma_k^x +\sin(\theta_k(\lambda))\, \id\right)
\ket{\downarrow_k},
\ee
where
\begin{equation}
\theta_k(\lambda)=\frac{1}{2}\arctan\left(\frac{\sin\left(\lambda\pi/2\right)\sin(k)}{\cos\left(\lambda\pi/2\right)+\sin\left(\lambda\pi/2\right)\cos(k)}\right).
\end{equation}
In the standard fermionic formalism, this reads:
\be
\ket{\psi(\lambda)}\equiv\prod_{k\in\mathcal{K}^+} \left(\cos(\theta_k(\lambda)) \, \opcdag{k} \opcdag{-k} + \sin(\theta_k(\lambda)) \right) \ket{0} \,,
\ee
where $\ket{0}$ is the vacuum state. Thanks to a Fourier transform, we can represent the ground states in position space as
\be
\ket{\psi(\lambda)}\equiv\prod_{k\in\mathcal{K}^+} \left(\frac{i}{N}\cos(\theta_k)\sum_{j,j'}e^{ik(j-j')} 
\opcdag{j} \opcdag{j'} + \sin(\theta_k)\right)\ket{0} \;.
\ee

The commutator matrix is related to the pseudospin commutator matrix 
\[
K'_{\{k,\alpha\}\{k',\alpha'\}}(\lambda)\equiv\bra{\psi(\lambda)}-i[\tilde\sigma_{k'}^{\alpha'},\tilde\sigma_k^\alpha]\ket{\psi(\lambda)}
\]
as follows:
\be
K^{(l)}_{\{n,\nu\}\{m,\mu\}}=\bra{\psi(\lambda)} -i[\Sigma_m^\mu,\Sigma_n^\nu]\ket{\psi(\lambda)}=
\sum_{k,k'}^{\mathcal{K}^+} \sum_{\alpha,\alpha'}^{x,y,z}
\mathcal{F}_{\{n,\nu\}\{k,\alpha\}}^{(l)}\mathcal{F}_{\{m,\mu\}\{k',\alpha'\}}^{(l)}K'_{\{k,\alpha\}\{k',\alpha'\}} \;,
\ee
where $\mathcal{F}_{\{n,\nu\}\{k,\alpha\}}^{(l)}$ is a matrix Fourier transform. In particular, for $l<N/2$ we have:
\be
\mathcal{F}^{(l)}=\frac{1}{\sqrt{N}}\begin{pmatrix}
0 & 0 & 1/\sqrt{2} & 0 & 0 & 1/\sqrt{2} & \dots \\
\sin(1 k_1) & 0 & 0 & \sin(1 k_2) & 0 & 0 & \dots\\
0 & -\sin(1 k_1) & 0 & 0 & -\sin(1 k_2) & 0 & \dots\\
0 & 0 & \cos(1 k_1) & 0 & 0 & \cos(1 k_2) & \dots\\
\dots & \dots & \dots & \dots & \dots & \dots & \dots \\
\sin(l k_1) & 0 & 0 & \sin(l k_2) & 0 & 0 & \dots\\
0 & -\sin(l k_1) & 0 & 0 & -\sin(l k_2) & 0 & \dots\\
0 & 0 & \cos(l k_1) & 0 & 0 & \cos(l k_2) & \dots
\end{pmatrix},
\ee
and for $l=N/2$:
\be
\mathcal{F}^{(N/2)}=\frac{1}{\sqrt{N}}\begin{pmatrix}
0 & 0 & 1/\sqrt{2} & 0 & 0 & 1/\sqrt{2} & \dots \\
-\sin(1 k_1) & 0 & 0 & -\sin(1 k_2) & 0 & 0 & \dots\\
0 & \sin(1 k_1) & 0 & 0 & \sin(1 k_2) & 0 & \dots\\
0 & 0 & \cos(1 k_1) & 0 & 0 & \cos(1 k_2) & \dots\\
\dots & \dots & \dots & \dots & \dots & \dots & \dots \\
\sin((\frac{N}{2}-1) k_1) & 0 & 0 & \sin((\frac{N}{2}-1) k_2) & 0 & 0 & \dots\\
0 & -\sin((\frac{N}{2}-1) k_1) & 0 & 0 & -\sin(\frac{N}{2} k_2) & 0 & \dots\\
0 & 0 & \cos((\frac{N}{2}-1) k_1) & 0 & 0 & \cos((\frac{N}{2}-1) k_2) & \dots \\
\sin(N k_1/2)/\sqrt{2} & 0 & 0 & \sin(N k_2/2)/\sqrt{2} & 0 & 0 & \dots\\
0 & -\sin(N k_1/2)/\sqrt{2} & 0 & 0 & -\sin(N k_2/2)/\sqrt{2} & 0 & \dots
\end{pmatrix}.
\ee

The last step consists of calculating the elements of the matrix $K'(\lambda)$. Thanks to the simple commutation rules between pseudospins, this is a block-diagonal matrix:
\be
K'(\lambda)=\begin{pmatrix}
K_{k_{1}}(\lambda) & 0 & \dots \\
0 & K_{k_{2}}(\lambda) & \dots \\
\dots & \dots & \dots \\
\end{pmatrix},
\ee
where the blocks are labelled by the elements of $\mathcal{K}^+$. Exploiting Eq.~\eqref{eq:gss}, we obtain the explicit form of the submatrices:
\be
K_k(\lambda)=\begin{pmatrix}
0 & v_k^z(\lambda) & 0\\
-v_k^z(\lambda) & 0 & v_k^x(\lambda)\\
0 & -v_k^x(\lambda) & 0
\end{pmatrix}.
\ee
Now, the commutator matrix can be easily calculated through the Fourier transform of $K'(\lambda)$:
\be
K^{(l)}(\lambda)=\big(\mathcal{F}^{(l)}\big)\big(K'(\lambda)\big)\big(\mathcal{F}^{(l)}\big)^{\mathrm{T}} \;.
\ee

\subsection{IQA with fermionic Gaussian states: relationship with spin systems}\label{app:jw}

Here we show that, thanks to the Jordan-Wigner transformations~\cite{IsingSantoro}, all the operators in $\mathcal{L}^{(l)}$ are equivalent to a set of spin strings spanning over a space of Hamiltonians that encloses several important physical models, such as the Ising model in transverse field~\cite{PFEUTY197079} and the XY model~\cite{LIEB1961407}. As a consequence, our results about Gaussian states can be extended to a large class of spin systems.

We consider the operators in $\mathcal{S}=\{S_0^z,S_1^x,S_1^y,S_1^z,\dots,S_{N/2-1}^x,S_{N/2-1}^y,S_{N/2-1}^z,S_{N/2}^x,S_{N/2}^y\}$ acting on a system of $N$ spins, where
\begin{align}
S_0^z&=-\frac{1}{2\sqrt{N}}\sum_j^N  \PauliSigma_j^z\nonumber\\
S_m^x&=\frac{1}{4\sqrt{N}}\sum_j^N\left( \PauliSigma_j^x\PauliSigma_{j+1}^z\dots\PauliSigma_{j+m-1}^z\PauliSigma_{j+m}^x-\PauliSigma_j^y\PauliSigma_{j+1}^z\dots\PauliSigma_{j+m-1}^z\PauliSigma_{j+m}^y\right)\nonumber\\
S_m^y&=\frac{1}{4\sqrt{N}}\sum_j^N\left( \PauliSigma_j^x\PauliSigma_{j+1}^z\dots\PauliSigma_{j+m-1}^z\PauliSigma_{j+m}^y+\PauliSigma_j^y\PauliSigma_{j+1}^z\dots\PauliSigma_{j+m-1}^z\PauliSigma_{j+m}^x\right)\nonumber\\
S_m^z&=\frac{1}{4\sqrt{N}}\sum_j^N\left( \PauliSigma_j^x\PauliSigma_{j+1}^z\dots\PauliSigma_{j+m-1}^z\PauliSigma_{j+m}^x+\PauliSigma_j^y\PauliSigma_{j+1}^z\dots\PauliSigma_{j+m-1}^z\PauliSigma_{j+m}^y\right),
\end{align}
where $N$ is even and the boundary conditions are periodic: $\PauliSigma_{j+N}^\mu=\PauliSigma_{j}^\mu$. 

Let us consider the parity operator $P\equiv \prod_n\PauliSigma_{n,z}$. $P$ commutes with each element of $\mathcal{L}$, so we can divide the whole Hilbert space into two parity sectors and write the $S_n^\mu$ in a block-diagonal form:
\be
S_n^\mu=
\begin{pmatrix}
S_{n,0}^\mu & 0 \\
0 & S_{n,1}^\mu 
\end{pmatrix},
\ee
where the subscript $p\in\{0,1\}$ in $S_{n,p}^\mu$ is referred to the parity of the Hilbert subspace on which the operator acts ($0$ even, $1$ odd).

We define the Jordan-Wigner transformations
\begin{align}
K_j&=\prod_{j'=1}^{j-1} 
(\id - 2 \opcdag{j'} \opc{j'}) \nonumber\\
	\PauliSigma_j^z&=
 (\id-2 \opcdag{j} \opc{j} ) \nonumber\\
	\PauliSigma_j^x&= K_j 
 ( \opcdag{j}+ \opc{j}) \nonumber\\
	\PauliSigma_j^y&=i\, K_j
 ( \opcdag{j}- \opc{j} ) \;,
\end{align}
where the $\opcdag{j}$'s and the $\opc{j}$'s are fermionic creation and annihilation operators in position space. After some algebra, the action of the basis elements in the even parity sector, which contains the ground states of the Hamiltonians spanned by $\mathcal{S}$, can be written as:
\begin{align}
S_{0,0}^z&=\Sigma_0^z\nonumber\\
S_{m,0}^x&=\Sigma_m^x\nonumber\\
S_{m,0}^y&=\Sigma_m^y\nonumber\\
S_{m,0}^z&=\Sigma_m^z,
\end{align}
with antiperiodic boundary conditions $\opc{N+m}\equiv -\opc{m}$ and $1\leq m\leq N/2$ .

\subsection{IQA with fermionic Gaussian states: non-projected IQA}\label{app:non_proj}

\begin{figure*}
    \includegraphics[height=70mm]{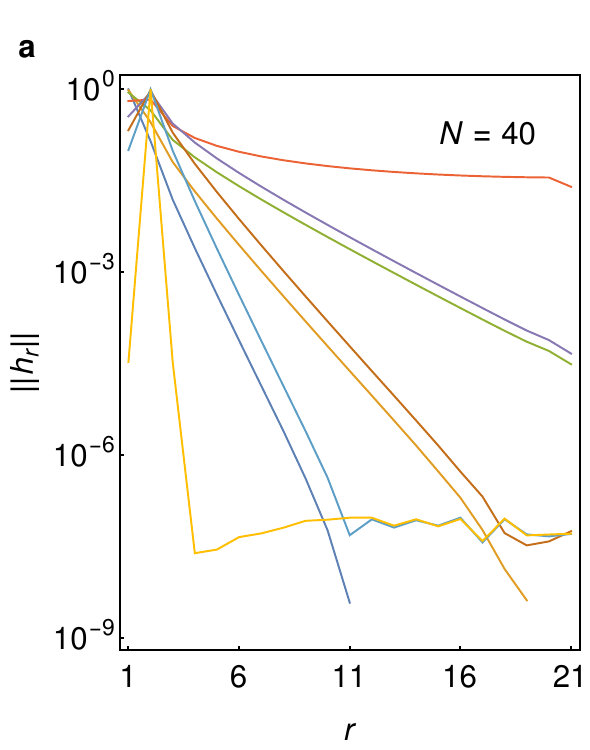}
    \includegraphics[height=70mm]{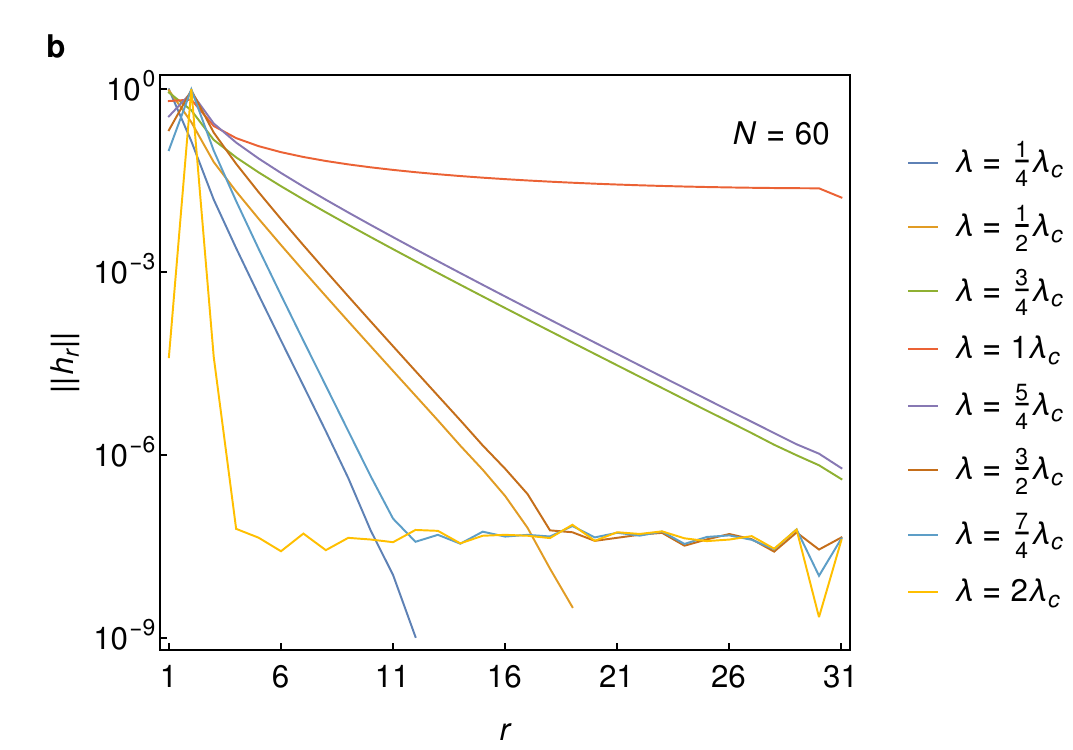}
\caption{IQA with fermionic Gaussian states --- Norm of the non-projected adiabatic Hamiltonian couplings of different ranges $r$. Panel (a): for $N=40$ sites, Panel (b): for $N=50$ sites}
\label{fig:weightsA}
\end{figure*}
\begin{figure*}
    \includegraphics[height=70mm]{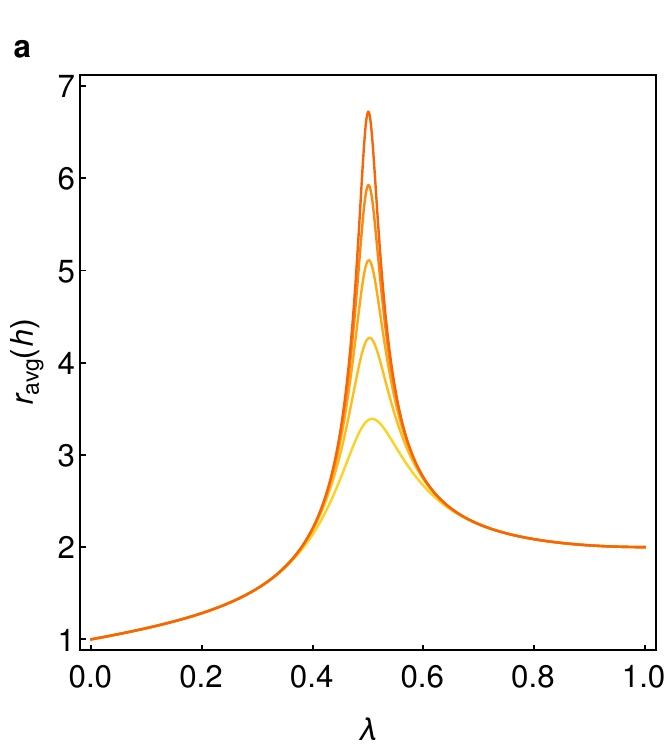}
    \includegraphics[height=70mm]{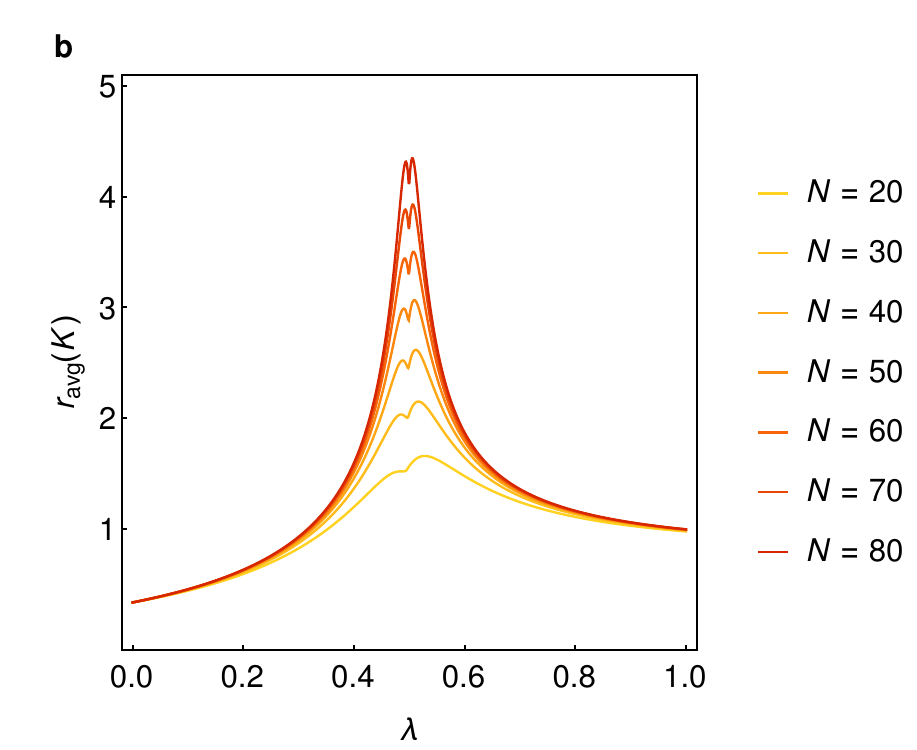}
\caption{IQA with fermionic Gaussian states --- Panel(a): effective interaction range of the non-projected adiabatic Hamiltonian. Panel (b): correlation range of $K_{ij}^{(N/2)}$.}
\label{fig:weightsB}
\end{figure*}

Here, we exploit Gaussian states to study the IQA for large system sizes without the TDVP. In this way, we test our claim that the non-projected dynamics of Eq.~(4)  generates $l$-local PHs for states with finite correlation length, making negligible the approximation induced by the TDVP.

We refer to the adiabatic solution to Eq.~(4) as \textit{non-projected} adiabatic Hamiltonian. Becouse of the definition of $\mathcal{L}^{(l)}$ for this many-body system, the evolution generated by $K_{\rmj,\rmj'}^{(N/2)}$ spans over the space of all translation-invariant quadratic fermionic Hamiltonians with antiperiodic boundary conditions, therefore it is equivalent to the non-projected evolution in Eq.~(4). We study the link between the (effective) interaction range of the non-projected Hamiltonian and the correlation length of $\ket{\psi(\lambda)}$, showing that the effective range of adiabatic Hamiltonian obtained through the non-projected dynamics scales with the correlation range of the ground state and therefore increases with the system size at the phase transition.

In Fig.~\ref{fig:weightsA}, we plot the total norm 
\beN
\lVert h_r\rVert=\sqrt{\sum_{i:\text{range}(L_i^{l})=r}h_i^2}
\eeN
of the couplings of range $r$ of this Hamiltonian at different values of $\lambda$, for a system of $40$ (a) and $60$ (b) sites. We observe that this norm exponentially decays with $r$ for $\lambda\neq\lambda_c = 1/2$ with a decay rate that does not depend on the system size. We can confirm that Eq.~(4) generates a local PH, but only for non-critical states.

We can define an effective interaction range as
\begin{equation*}
r_\text{avg}(h)\equiv \frac{\sum_r r \lVert h_r\rVert}{\sum_r \lVert h_r\rVert}.
\end{equation*}
This range is represented in Fig.~\ref{fig:weightsB}(a). We can see that, for sufficiently large $N$, it does not depend on the system size for non-critical states. However, it diverges linearly with $N$ at $\lambda=\lambda_c$. This scaling is reminiscent of $\ket{\psi(\lambda)}$ correlation length.

To understand how the scaling behavior originated in the equation of motion, we look at the commutator matrix $K_{\rmj,\rmj'}^{(N/2)}$. We define a correlation range for this matrix
\beN
r_\text{avg}(K)\equiv \frac{\sum_{\rmj,\rmj'} 
\lvert \rmj-\rmj'\rvert \; \lvert K_{\rmj,\rmj'}\rvert}{\sum_{\rmj,\rmj'} \lvert K_{\rmj,\rmj'} \rvert} \,,
\eeN
which measures the decay of non-diagonal elements. It is depicted in Fig.~\ref{fig:weightsB}(b), and follows the same scaling behavior as the previously analyzed functions. We can conclude that, when the correlation length of states is finite, the non-projected dynamics weakly couples the local and non-local operators and the Hamiltonian does not delocalize, and so the truncation at finite $l$ in Eq.~(6) is a good approximation far from the critical point.

\section{IQA with a non-integrable Ising chain ($B_z=0.8$): robustness against statistical errors}\label{app:robustness}

Any application of our method to Hamiltonian learning will introduce a statistical error scaling as $1/\sqrt{N_m}$ on the matrix $K_{\rmj,\rmj'}$, where $N_m$ is the total number of measurements~\cite{RATTACASO2022_HIGH}. For this reason, we have tested the effect of a statistical error on the local observables $K_{\rmj,\rmj'}$ on the final fidelity of the annealing process by adding to each matrix element a Gaussian noise with zero mean and different variances. As we observe in Figure~\ref{fig:gaussian_noise}, the adiabatic fidelity for a system of size $N=10$ respectively decreases to $F=0.998\pm 0.001$, $F=0.993\pm 0.007$ and $F=0.96\pm 0.02$ in correspondence to the variances $\sigma\in\{0.01 K_{\rmj,\rmj'}, 0.02 K_{\rmj,\rmj'}, 0.05 K_{\rmj,\rmj'}\}$, demonstrating the robustness of our method.

\begin{figure*}[ht!]
    \includegraphics[height=5.5cm]{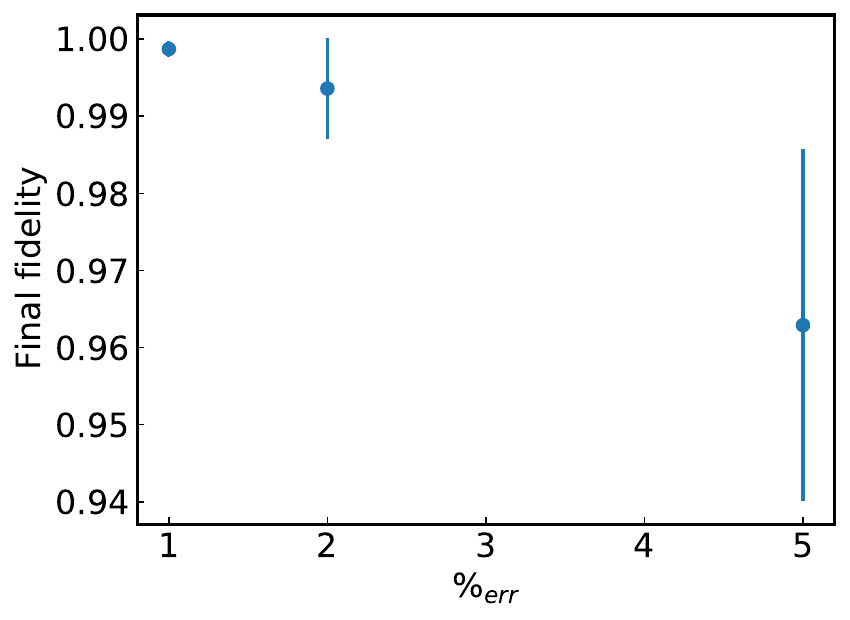} 
  \caption{IQA with a non-integrable Ising chain  ($B_z=0.9$) --- Final fidelity of the IQA at $T=1000$ for a system of $10$ spins as a function of the relative statistical error on the matrix elements $K_{\rmj,\rmj'}$, simulated as a Gaussian noise.}
    \label{fig:gaussian_noise}
\end{figure*}

\subsection{IQA with a non-integrable Ising chain - intermediate annealing time behavior}\label{app:additional}

When the IQA is applied to the non-integrable Ising chain examined in the main text, some intermediate annealing times are characterized by local drops in fidelity. As shown in Figure~\ref{fig:non-integrable-additional} (a) and (b), these drops occur for both $B_z=0.8$ and $B_z=0.9$. Subsequently, the fidelity definitively converges to one. This convergence of fidelity corresponds to the adiabatic limit, as seen in Figure~\ref{fig:non-integrable-additional} (c) and (d), which shows the relative distance $R_{T,\Delta T}$ between solutions with different annealing times.

In this analysis, we focus on the annealing process near a fidelity drop, specifically at $B_z=0.9$ and $T=512$, for a system of $8$ spins. We examine the fidelity between the target state $\ket{\psi(\lambda)}$ and the ground state $\ket{\psi_0}$ of the IQA Hamiltonian, as well as the first excited state $\ket{\psi_1}$, at three different annealing times: before the drop ($T=256$), at the drop ($T=512$), and after the drop ($T=8192$). As depicted in Figure~\ref{fig:non-integrable-additional}, we observe that the target state oscillates between the ground state and the first excited state of the adiabatic Hamiltonian, with a crossing occurring at $T=512$. This crossing corresponds to the closure of the first energy gap of the Hamiltonian. This transient behavior vanishes for longer times and is reminiscent of Landau-Zener oscillations, which also characterize standard quantum annealing.

\begin{figure*}[h]
    \includegraphics[height=5.3cm]{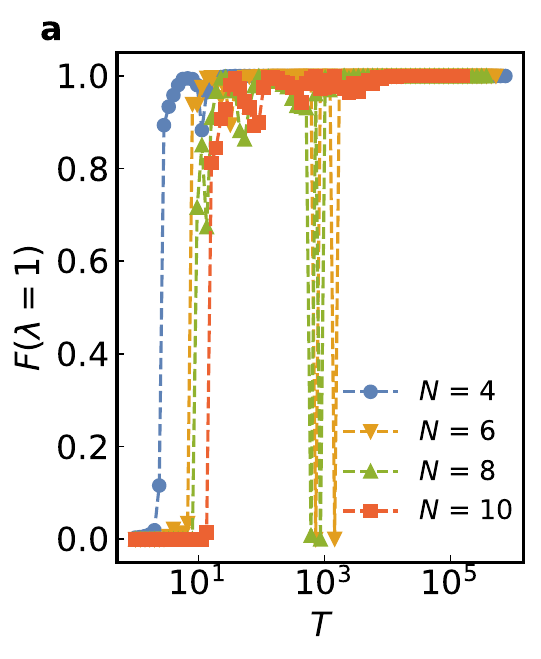} 
    \includegraphics[height=5.3cm]{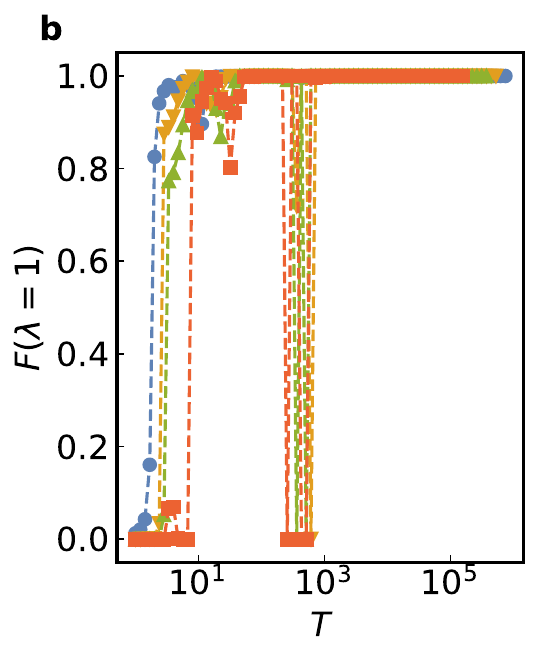}
    \includegraphics[height=5.3cm]{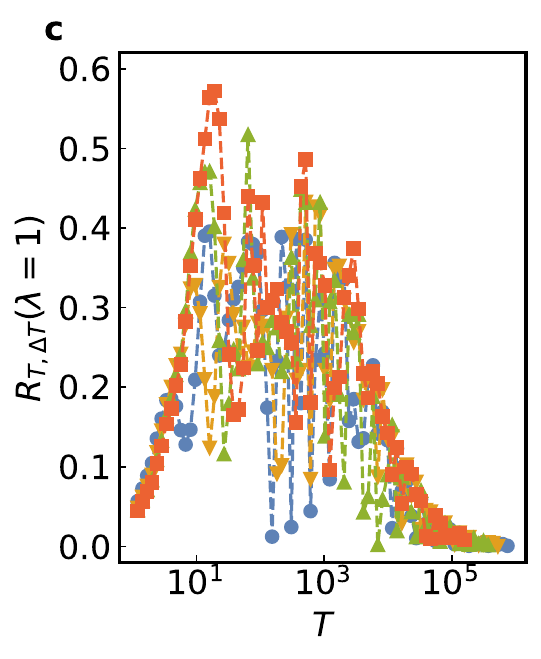} 
    \includegraphics[height=5.3cm]{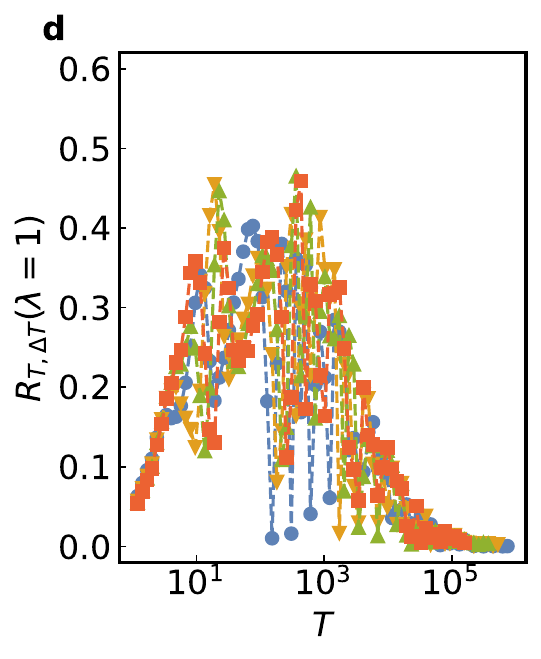}
\caption{IQA with a non-integrable Ising chain --- In Panels (a) and (b), final fidelity between the target state $\ket{\psi(\lambda=1)}$ and the ground state $\ket {\psi_\text{GS}^{(l)}(\lambda=1)}$ of the adiabatic $2$-local Hamiltonian, for different system sizes and different annealing times, for $B_z=0.8$ in Panel (a) and for $B_z=0.9$ in Panel (b). In Panels (c) and (d), the final value of the relative distance $R_{T,\Delta T}(\lambda=1)$ between solutions of Eq.~(6) with different annealing times $T$ and $T+\Delta T = 2T$, as a function of the annealing time, for systems of different sizes. In Panel (c) for $B_z=0.8$, in Panel (d) for $B_z=0.9$.}
    \label{fig:non-integrable-additional}
\end{figure*}

\begin{figure*}[h]
    \includegraphics[height=5.4cm]{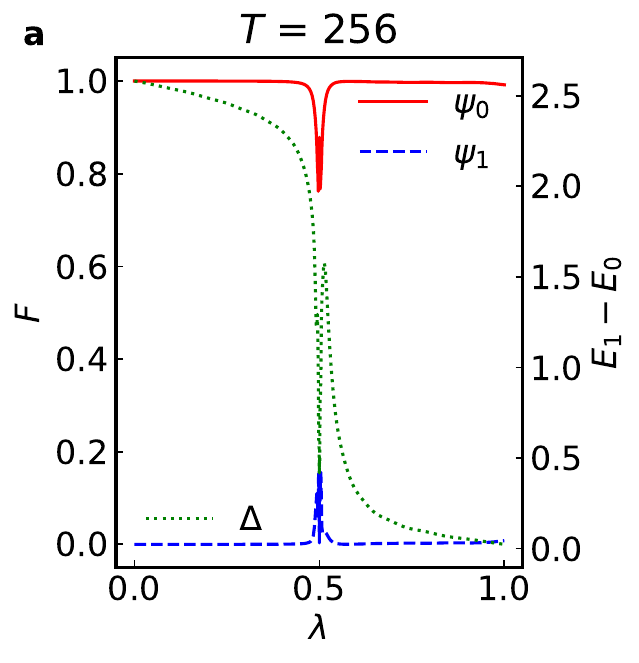} 
    \includegraphics[height=5.4cm]{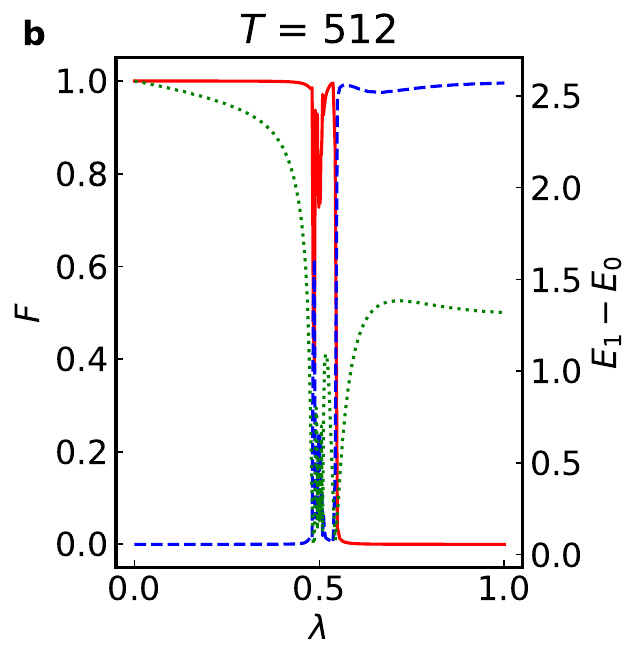}
    \includegraphics[height=5.4cm]{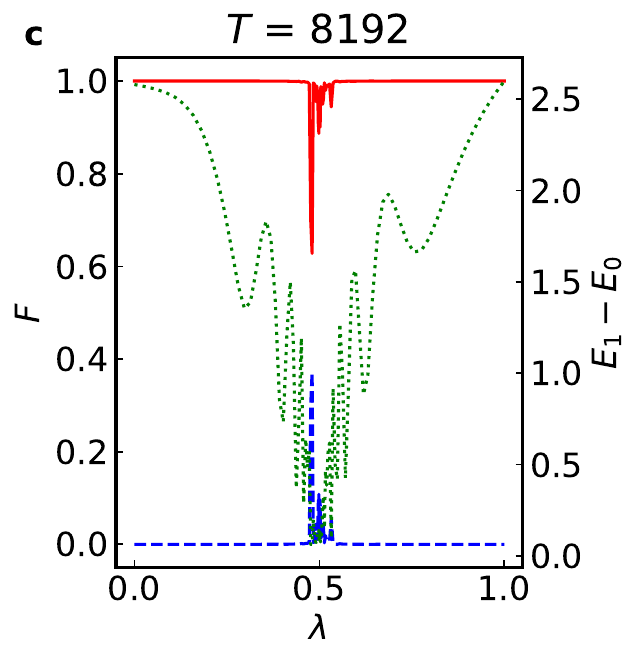}
\caption{Fidelity between the target state and the ground state $\ket{\psi_0}$ of the IQA Hamiltonian, fidelity between the target state and the first excited state $\ket{\psi_1}$ of the IQA Hamiltonian and first energy gap of the IQA Hamiltonian as a function of $\lambda$, at $B_z=0.9$ for a system of $8$ spins. In Panel (a) $T=256$, in Panel (b) $T=512$, in Panel (c) $T=8192$.}
  \label{fig:drop_study}
\end{figure*}

\end{adjustwidth}

\end{document}